\documentclass[epsfig, 12pt]{article}
\usepackage{graphicx}
\usepackage{epstopdf}

\usepackage{array}
\usepackage{amsmath}
\usepackage{amssymb}

\newcommand{\beq}{\begin{equation}}   
\newcommand{\eeq}{\end{equation}}
\newcommand{\beqn}{\begin{eqnarray}}   
\newcommand{\eeqn}{\end{eqnarray}}

\begin{document}

\unitlength = 1mm

\def\de{\partial}
\def\Tr{ \hbox{\rm Tr}}
\def\const{\hbox {\rm const.}}  
\def\o{\over}
\def\im{\hbox{\rm Im}}
\def\re{\hbox{\rm Re}}
\def\bra{\langle}\def\ket{\rangle}
\def\Arg{\hbox {\rm Arg}}
\def\Re{\hbox {\rm Re}}
\def\Im{\hbox {\rm Im}}
\def\diag{\hbox{\rm diag}}


\def\QATOPD#1#2#3#4{{#3 \atopwithdelims#1#2 #4}}
\def\stackunder#1#2{\mathrel{\mathop{#2}\limits_{#1}}}
\def\stackreb#1#2{\mathrel{\mathop{#2}\limits_{#1}}}
\def\Tr{{\rm Tr}}
\def\res{{\rm res}}
\def\Bf#1{\mbox{\boldmath $#1$}}
\def\balpha{{\Bf\alpha}}
\def\bbeta{{\Bf\beta}}
\def\bgamma{{\Bf\gamma}}
\def\bnu{{\Bf\nu}}
\def\bmu{{\Bf\mu}}
\def\bphi{{\Bf\phi}}
\def\bPhi{{\Bf\Phi}}
\def\bomega{{\Bf\omega}}
\def\blambda{{\Bf\lambda}}
\def\brho{{\Bf\rho}}
\def\bsigma{{\bfit\sigma}}
\def\bxi{{\Bf\xi}}
\def\bbeta{{\Bf\eta}}
\def\d{\partial}
\def\der#1#2{\frac{\d{#1}}{\d{#2}}}
\def\Im{{\rm Im}}
\def\Re{{\rm Re}}
\def\rank{{\rm rank}}
\def\diag{{\rm diag}}
\def\2{{1\over 2}}
\def\ntwo{${\mathcal N}=2\;$}
\def\nfour{${\mathcal N}=4\;$}
\def\none{${\mathcal N}=1\;$}
\def\ntwot{${\mathcal N}=(2,2)\;$}
\def\ntwoo{${\mathcal N}=(0,2)\;$}
\def\x{\stackrel{\otimes}{,}}

\newcommand{\cpn}{CP$(N-1)\;$}
\newcommand{\wcpn}{wCP$_{N,\tilde{N}}(N_f-1)\;$}
\newcommand{\wcpd}{wCP$_{\tilde{N},N}(N_f-1)\;$}
\newcommand{\vp}{\varphi}
\newcommand{\pt}{\partial}
\newcommand{\ve}{\varepsilon}
\renewcommand{\theequation}{\thesection.\arabic{equation}}

\newcommand{\sun}{SU$(N)\;$}

\vfill

\begin{titlepage}

\begin{flushright}
FTPI-MINN-16/08, UMN-TH-3519/16
\end{flushright}

\vspace{1mm}

\begin{center}
{  \Large \bf  
 Heterotic Non-Abelian String of a Finite \\[2mm]
 Length 
 }

\vspace{5mm}

 {\large \bf  S.~Monin$^{\,a}$,  M.~Shifman$^{\,a}$ and \bf A.~Yung$^{\,\,a,b,c}$}
\end {center}

\begin{center}

$^a${\it  William I. Fine Theoretical Physics Institute,
University of Minnesota,
Minneapolis, MN 55455, USA}\\
$^{b}${\it National Research Center ``Kurchatov Institute'', 
Petersburg Nuclear Physics Institute, Gatchina, St. Petersburg
188300, Russia}\\
$^{c}${\it  St. Petersburg State University,
 7/9 Universitetskaya nab., St. Petersburg 199034, Russia}
\end{center}

\vspace{4mm}

\begin{center}
{\large\bf Abstract}
\end{center}

We consider non-Abelian strings in \ntwo supersymmetric QCD with 
the U$(N)$ gauge group and $N_f=N$
quark flavors deformed by a mass term for the adjoint matter. This deformation breaks \ntwo
supersymmetry down to \none\!. Dynamics of orientational zero modes on the string world sheet
are described then by  CP$(N-1)$ model with  \ntwoo \! supersymmetry.
We study the string of a finite length $L$ assuming  compactification   
on a cylinder (periodic boundary conditions). The world-sheet theory is solved in the large-$N$ 
approximation.  We  find
a rich phase structure in the  $(L, \,u)$ plane where $u$ is a deformation parameter.
 At large  $L$  and intermediate $u$ we find 
  a  phase  with broken $Z_{2N}$ symmetry, $N$ vacua and a
 mass gap.  At large  values of $L$  and  $u$ still larger we have the $Z_{2N}$-symmetric phase
with a single vacuum and massless fermions.
In both phases  \ntwoo supersymmetry is spontaneously broken.
 We also observe a phase with broken \sun symmetry at small $L$.
 In the latter phase the mass gap vanishes and the vacuum energy is zero 
 in the leading $1/N$ approximation.
However, we expect that $1/N$
 corrections will break \ntwoo supersymmetry. We also discuss how this rich phase 
 structure matches  
the \ntwot limit in which the world-sheet theory has a single phase with the mass
gap independent of $L$.

\vspace{2cm}

\end{titlepage}

\section{Introduction}
\setcounter{equation}{0}

Recently there was a considerable progress in the study of long confining strings of a fixed length 
both on lattices \cite{latt,Athen} and by constructing  the effective theory on the string world sheet,
see \cite{AharonyKomar,Dubovsky}. In our recent paper \cite{MSY15} we initiated a study of 
a closed non-Abelian 
string of a finite length $L$ assuming  compactification   
on a cylinder with circumference $L$ (periodic boundary conditions).

Non-Abelian strings were first found in \ntwo supersymmetric gauge theories 
\cite{HT1,ABEKY,SYmon,HT2}. Later
this construction was generalized to a wide class of non-Abelian gauge theories,
both supersymmetric
and non-supersymmetric, see \cite{Trev,Jrev,SYrev,Trev2}.
Both Abelian and non-Abelian strings have translational modes associated with broken 
translation symmetries. The main feature of the non-Abelian strings is the 
occurrence of extra moduli:
orienational zero modes associated with the color flux rotation in the internal
space. Dynamics 
of these orientational moduli is described by two-dimensional CP$(N-1)$ model on
the string world-sheet. The translational modes are completely decoupled.

In \cite{MSY15} we  studied both non-supersymmetric case as well as 
1/2-BPS string in \ntwo supersymmetric QCD. For non-supersymmetric case we found a phase transition 
in the world-sheet  theory in the large-$N$ limit. At large $L$ this 
theory develops a mass gap and  is in the Coulomb/confinement 
phase. Finite-length effects are exponentially suppressed. 
At small lengths it is in the deconfinement
phase.  

 \ntwo supersymmetric QCD has 
eight supercharges and, since our strings are 1/2-BPS, the world-sheet CP$(N-1)$ model
has \ntwot supersymmetry.
In this  case we found a single phase with a mass gap and 
unbroken supersymmetry  \cite{MSY15}. The mass gap
turns out to be independent of the  string length.

If we introduce a mass term for the adjoint matter in the bulk we break bulk \ntwo 
supersymmetry down to \none\!\!. The string remains BPS saturated  \cite{SY05}.
It was conjectured by  Edalati and Tong  
\cite{ET07} and confirmed in \cite{SYhet} that the target space in the deformed model is 
$CP(N-1)\times C$. The right-handed supertranslational modes become coupled to 
superorientational ones,  and the world sheet theory becomes  heterotic  model
with \ntwoo supersymmetry. It is important that this is a nonminimal model (cf. \cite{op}) well defined for all $N$. 

\begin{figure}[h!]
\begin{center}
\includegraphics[width=11cm]{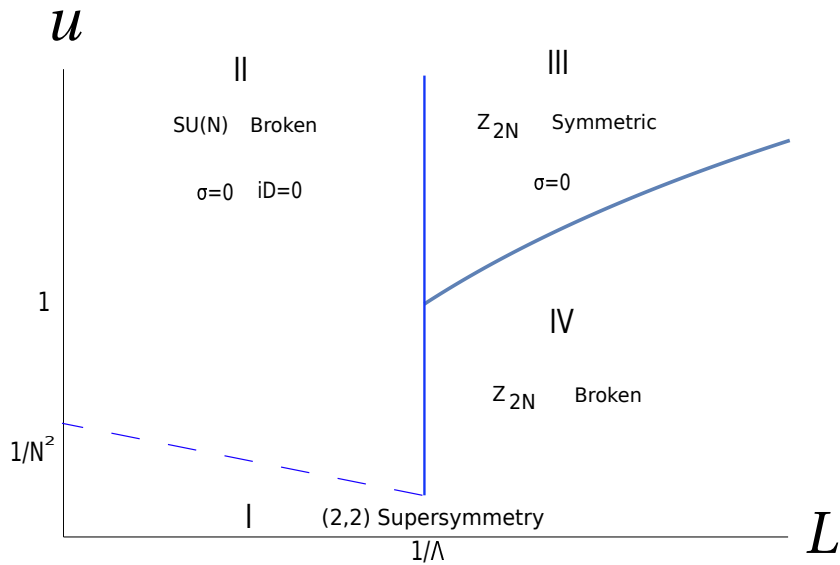}
\end{center}
\label{phasediag}
\caption{(\mbox{I}) $u<1/N^2$ region corresponds to the \ntwot solution regardless of 
$L$; (\mbox{II}) $u\gg1/N^2$ and $L<1/\Lambda$ region corresponds to \sun-
broken phase ($n^l$ fields develop VEV). The line separating these two phases 
is dashed since we don't know its exact form; (\mbox{III}) $L>1/\Lambda$ and large $u$ region
represents the $Z_{2N}$-symmetric phase with massless fermions; (\mbox{IV}) $L>1/\Lambda$ and 
moderate $u$ region represents $Z_{2N}$-broken phase with massive bosons and fermions.}
\end{figure}

In this paper we solve the above heterotic \ntwoo \cpn model  on  a cylinder with 
circumference $L$ in the large-$N$ approximation, assuming periodic boundary conditions.
Our solution is drastically different from the one obtained in the \ntwot case. First of all we observe 
three distinct phases instead of one. Two phases (III and IV in Fig. \ref{phasediag}) preserve the \sun 
global symmetry. The finite-$L$ effects are exponentially suppressed at large $L$ and intermediate
 values of the deformation parameter $u$, in much the same way as 
in non-supersymmetric theory \cite{MSY15}. The parameter of deformation $u$ is 
related to the mass of the adjoint
field in the bulk SQCD. The theory in this phase has mass gap and $N$ vacua; 
the discrete chiral  $Z_{2N}$
symmetry is spontaneously broken down to $Z_2$.

 As we increase $u$ still keeping    $L$ large the theory undergoes a  third order phase transition into
a phase with a single vacuum and unbroken $Z_{2N}$. This is a  phase with massless fermions.
 A sketch of the full phase diagram of the world sheet theory in the  $(L,\,u)$ plane is shown in 
 Fig.~\ref{phasediag}.

As was the case for non-supersymmetric theory,  we  find a phase with broken \sun symmetry
at small $L$.
In the latter phase mass gap  is  zero in the leading 
approximation. Moreover, we find that the vacuum energy also vanishes at $N=\infty$. However, we expect corrections
of higher order in $1/N$  (or, perhaps, exponential corrections
$e^{-N}$) to   break  \ntwoo supersymmetry  and lift
the vacuum  energy. 

We discuss how this rich phase structure evolves to the ${\mathcal N} =(2,2)$ picture with a single phase
 in the  limit of zero deformation, $u=0$.
 
 In the \ntwot  problem supersymmetry is unbroken and we deal
with a single phase with an  $L$ independent mass gap (the latter property is dictated by holomorphy \cite{MSY15}).
The limit $u\to 0$ turns out to be rather subtle. It turns out that the relevant parameter which 
ensures the saddle point approximation  used in the large-$N$ method is 
$uN^2$ rather than $N$. If the deformation parameter $u\sim 1$ the large-$N$ limit ensures the validity of the quasiclassical
approximation in the effective one-loop action. However, at extremely small $u$ 
 this approximation breaks down. To get a smooth  $u\to 0$ limit
we quantize  the  holonomy of 
the two-dimensional gauge potential around the compact spatial dimension of the string. The  Polyakov line
\beq
\exp\left(i\int dx_k A_k\right)
\label{Polyakovloop}
\eeq
 depends  only on time. Hence  we 
consider a quantum-mechanical problem 
averaging this operator over the appropriate wave functions. This gives us the desired smooth $u\to 0$
limit and we  recover the \ntwot result in the narrow strip $u\sim 1/N^2$, see  Fig.~\ref{phasediag}.

The paper is organized as follows. In Sec. \ref{infinitevol} we review the large-$N$ solution of 
the heterotic \cpn model on the infinite two-dimensional plane. In Sec. \ref{equations} we present
the large-$N$ solution for the heterotic string compactified on a cylinder. In Sec. \ref{zNbroken}
 we discuss the \sun\!\!-symmetric phase with broken $Z_{2N}$ symmetry while in Sec. \ref{zNunbroken}
we consider the $Z_{2N}$ unbroken phase.  In 
 Sec. \ref{brphase} we deal with the \sun broken phase. In Sec. \ref{qmsmallL}
we discuss the $u\to 0$ limit at small $L$. Section \ref{conclusions} summarizes
our conclusions.

\section{Heterotic \boldmath\ntwoo \cpn model at $L=\infty$}
\setcounter{equation}{0}
\label{infinitevol}

The heterotic \ntwoo \cpn model at $L=\infty$ was solved  in \cite{SY08}  in the large-$N$ limit. 
In this section we will
briefly review this solution. The bosonic part of the action in the gauged 
formulation is 
\beq
S_b=\int d^2x \Big[|\nabla_kn^l|^2+2|\sigma|^2|n^l|^2+iD(|n^l|^2-r_0)
+4|\omega|^2|\sigma|^2\Big]\,,
\label{Sb}
\eeq
where $n^l$ ($l=1,...N$) is a complex $N$-vector parametrizing the 
orientational modes.
Moreover,
$$\nabla_k=\partial_k-iA_k\,.$$ Here $A_k$ is the  gauge potential, $\sigma$ is a complex scalar field.
The fields  $A_k$, $\sigma$ and $D$ belong to the gauge (vector) multiplet. These fields come
without kinetic terms and are auxiliary.  Moreover, $r_0$ 
is a coupling constant,  while $\omega$
is the $(2,2)$ deformation parameter. 
Eliminating $D$ leads to the constraint 
\beq
|n^l|^2=r_0\,.
\eeq
The fermionic part of the action is  
\beqn
S_f&=&\int d^2x \Big[\bar\xi_{lR}i(\nabla_0-i\nabla_3)\xi_R^l+\bar
\xi_{lL}i(\nabla_0+i\nabla_3)\xi_L^l\nonumber \\
&+&i\sqrt{2}\sigma\bar\xi_{lR}\xi_L^l+i\sqrt{2}\bar{n}^l(\lambda_R\xi^l_L-\lambda_
L\xi^l_R)\nonumber \\[3mm]
&+&i\sqrt{2}\sigma^\star\bar\xi_{lL}\xi_R^l+i\sqrt{2}(\bar{\lambda}_L\bar{\xi}^l_R
-\bar{\lambda}_R\bar{\xi}^l_L)n^l\nonumber \\[3mm]
&+&\frac{1}{2}\bar\zeta_Ri\partial_L\zeta_R+(i\sqrt{2}\omega\bar\lambda_L\zeta_R
+\mbox{H.c.})\Big]\,,
\label{Sf}
\eeqn
where $\xi^l_{R,L}$ are fermionic superpartners of $n^l$ (superorientational modes of the string),
$\lambda_{R,L}$ are auxiliary fermions from the vector superfield, while $\zeta_R$ is the right-handed
supertranslational mode. In the (2.2) model it was decoupled. We do not include the bosonic translational modes 
describing shifts of the string center. Nor do we include 
the left-handed supertranslational mode  $\zeta_L$,    because both decouple not only in the (2.2) but in the (0.2) model
as well \cite{ET07,SYhet}.

The terms containing $\zeta_R$ or $\omega$ break \ntwot down to \ntwoo. The 
deformation parameter $\omega$ is complex and scales with $N$ as \cite{SY08}
\beq
\omega\sim\sqrt{N}\,.
\eeq
It is determined by the mass parameter of the adjoint matter in the bulk theory \cite{SYhet}.

Integrating over $\lambda_{L,R}$  leads to the constraints
\beqn
\bar{n}^l\xi^l_L&=&0\nonumber\,, \\[2mm]
\bar\xi_R n^l&=&\omega\zeta_R\,.
\label{orth}
\eeqn
Integrating over  $\sigma$  implies
\beq
\sigma=-\frac{i}{\sqrt{2}(r_0+2|\omega|^2)}\bar\xi_{lL}\xi_R^l\,.
\label{chiral}
\eeq
Note that this model has an axial $U(1)$ symmetry 
broken by the chiral anomaly down to $Z_{2N}$ much in the same way as in 
the \ntwot model \cite{W79}. We find that $\sigma$ develops a vacuum expectation value (VEV) which
results in a spontaneous breaking of the discrete $Z_{2N}$ down to $Z_2$. Moreover
as can be seen from (\ref{chiral}), a non-zero VEV of the $\sigma$ field corresponds to a non-zero
fermion bilinear condensate $\left<\bar\xi_{lL}\xi_R^l\right>$.

\vspace{1mm}

Since both fields $n^l$ and $\xi^l$ appear in the action quadratically we can integrate them out.
This produces the product of two determinants,
\beq
\mbox{det}^{-N}\left(-\partial_i^2+iD+2|\sigma|^2\right)\mbox{det}^N\left(-\partial_i^2+2|\sigma|^2\right).
\label{sdetzero}
\eeq
The first determinant comes from the boson $n^l$ fields, while the second comes from the fermion
 $\xi^l$ fields. 
Note 
that if $D=0$ the two contributions obviously cancel each other, and supersymmetry is 
unbroken. Also, the non-zero values of $iD+2|\sigma|^2$ and $2|\sigma|^2$ can be interpreted as non-zero 
values of the masses of the $n^l$ and $\xi^l$ fields, respectively.
We put $A_k=0$. 

The final expression for the effective potential is  (see 
\cite{SY08})
\beqn
V_{\rm eff}&=&\int d^2x\frac{N}{4\pi}\Bigg[-(iD+2|\sigma|^2)\ln\frac{iD+2|\sigma|^2}{\Lambda^2}
+iD\nonumber \\
&+&2|\sigma|^2\ln\frac{2|\sigma|^2}{\Lambda^2}+2|\sigma|^2u\Bigg],
\label{infLV}
\eeqn
where the logarithmic ultraviolet divergence of the coupling constant is traded for 
the finite scale $\Lambda$ of the asymptotically free \cpn model. We also introduced a dimensionless 
deformation parameter 
\beq
u=\frac{8\pi}{N}|\omega|^2\,,
\label{29}
\eeq
which does not scale with $N$. 

To find the saddle point we minimize the potential 
with respect to $D$ and $\sigma$, which yields the following set of equations:
\beqn
\ln\frac{iD+2|\sigma|^2}{\Lambda_{CP}^2}=0\,,\nonumber \\[3mm]
\ln\frac{iD+2|\sigma|^2}{2|\sigma|^2}=u\,.
\label{infLeqs}
\eeqn
The solution to these equations is
\beq
iD=\Lambda^2(1-e^{-u})\,, \quad \mbox{and}\; \quad 2|\sigma|^2=\Lambda^2e^{-u}\,.
\eeq
The value of $D$ in this solution does not vanish, implying that supersymmetry is spontaneously broken.
We see that $\sigma$ develops a VEV giving masses to the $n^l$ fields and their 
fermion superpartners $\xi^l$. More exactly, the
solution for $\sigma$ can also be written as
\beq
\sqrt{2}\sigma=\Lambda \exp\left(-\frac{u}{2}+\frac{2\pi i k}{N}\right)\,, 
\quad k=0,...,N-1\,,
\label{sigmavev}
\eeq
where the phase factor is not seen in Eq. (\ref{infLeqs}). It comes as a result of a chiral anomaly
which breaks the chiral U(1) symmetry,
U$(1) \to Z_{2N}$. The field  $\sigma$ has the chiral charge 2. Thus a non-zero VEV of $|\sigma|$
ensures that $Z_{2N}$ symmetry is broken down to $Z_2$ and there are $N$ 
vacua presented in (\ref{sigmavev}). 

Substituting the solution (\ref{infLeqs}) into (\ref{infLV}) we obtain an expression for 
the vacuum energy density
\beq
V_{\mbox{vac}}=\frac{N}{4\pi}\Lambda^2(1-e^{-u})\,,
\eeq
which, as expected,  vanishes in the limit $u\rightarrow0$\,.

\section{\boldmath\ntwoo model on a cylinder }
\setcounter{equation}{0}
\label{equations}

The \ntwot model on a cylinder was solved in the large-$N$ limit in \cite{MSY15}.
In this section we apply the same approach to \ntwoo model assuming periodic boundary conditions
both for bosons and fermions.
Since the action (\ref{Sb}) and (\ref{Sf}) is quadratic in $n^l$ and $\xi^l$ we 
can integrate over these fields. We assume that the compact dimension in the bulk theory
is $x_1$ and the string is wrapped around this dimension. We will assume a nontrivial holonomy
(\ref{Polyakovloop}) of $A_k$ 
around this compact dimension. In  the $A_0=0$ gauge we will look for a solution with $\,A_1=\mbox{const}$.

First consider the case when neither of the fields $n^l$ or $\xi^l$ develop VEVs. 
The expression for the effective potential is easily found,
\beqn
V&=&\frac{N}{4\pi}\Big(iD-iD\ln\frac{m_b^2}{\Lambda^2}-m_f^2\ln\frac{m_b^2}{m_f^2}
+m_f^2u\nonumber \\[2mm]
&+&8m_f^2\sum_{k=1}^\infty\frac{K_1(Lm_fk)}{Lm_fk}\cos LkA_1
\nonumber \\[2mm]
&-&
8m_b^2\sum_{k=1}^\infty\frac{K_1(Lm_bk)}{Lm_bk}\cos LkA_1\Big)\,,
\label{Vs}
\eeqn
where we use an effective mass notation for the bosonic $n^l$ and fermionic $\xi^l$ fields, 
\beq
 m_b^2=iD+2|\sigma|^2, \qquad m_f^2=2|\sigma|^2,\,.
\label{mb,f}
\eeq
Here  $K_1(z)$ is the modified
Bessel function of the second kind and the deformation parameter $u$ is related to the  
parameter $\omega$ as in (\ref{29}).
The first line in (\ref{Vs}) is the same as the one found in the case of the $L=\infty$
string (\ref{infLV}), while the second and third lines represent contributions arising 
due to the finite length of the string. The potential (\ref{Vs}) is periodic in the phase $LA_1$, with the
period $2\pi$, so we can assume that $0\le LA_1 < 2\pi$.

\subsection{Saddle point approximation}

To find VEVs of $A_1,\,$ of 
$\sigma\,$ and $iD$ we take derivatives of (\ref{Vs}) with respect to these fields.
 Then we obtain three equations, 
\beqn
V_{N, A_1}&=&m_b\sum_{k=1}^\infty K_1(Lm_bk)\sin LkA_1
-m_f\sum_{k=1}^\infty K_1(Lm_fk)\sin LkA_1 \,,\nonumber \\[2mm]
V_{N, \sigma^\star}&=&2\sigma\left[-\ln\frac{m_b^2}{m_f^2}+4\sum_{k=1}^\infty 
K_0(Lm_bk)\cos LkA_1\right.
\nonumber \\[2mm]
&-&
\left. 4\sum_{k=1}^\infty K_0(Lm_fk)\cos LkA_1+u\right],\nonumber \\[2mm]
V_{N, iD}&=&-\ln\frac{m_b^2}{\Lambda^2}+4\sum_{k=1}^\infty K_0(Lm_bk)\cos LkA_1\,.
\label{SPs}
\eeqn
One can see that the first equation is satisfied when either $A_1=0$ or $A_1=\pi/L$. 
However, unlike the bosonic theory \cite{MSY15}, $A_1=0$ corresponds to the maximum of potential.
The energy is lower if $LA_1=\pi$. This can be easily understood. Consider the second and third lines
in (\ref{Vs}),
\beq
V_A\sim \left[m_fK_1(Lm_f)-m_bK_1(Lm_b)\right]\cos(LA_1)\,.
\eeq
On the one hand we know from the definition that $m_b\geq m_f$. On the other 
hand $K_1(x)$ decreases exponentially at large values of the argument. 
Thus, at least for large $L$ the potential $E_A=c\times \cos(LA_1)$, where $c>0$.
Hence we conclude that the minimum of the potential is at $LA_1=\pi$. This 
conclusion is also supported by a numerical calculation, see Figs. 2,3. Below  we assume that
\beq
LA_1=\pi\,.
\label{Api}
\eeq
 
\begin{figure}[h!]
\begin{center}
\includegraphics[width=7.5cm]{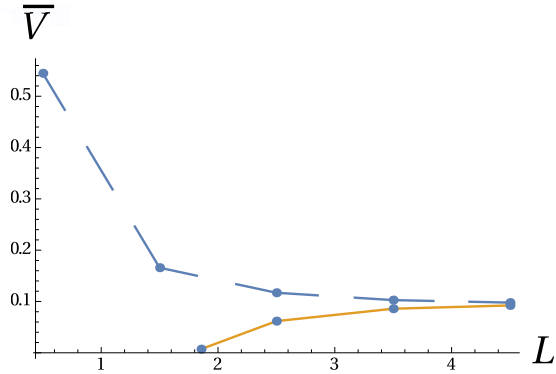}
\end{center}
\caption{\small $\bar{V}\equiv4\pi V$ vs string length $L$ at the value of deformation parameter $u=0.1$.
Solid line corresponds to $A_1=\pi/L$, while dashed line correcponds to $A_1=0$.}
 \label{V-L}
 \end{figure}
\begin{figure}[h!]
\begin{center}
\includegraphics[width=7.5cm]{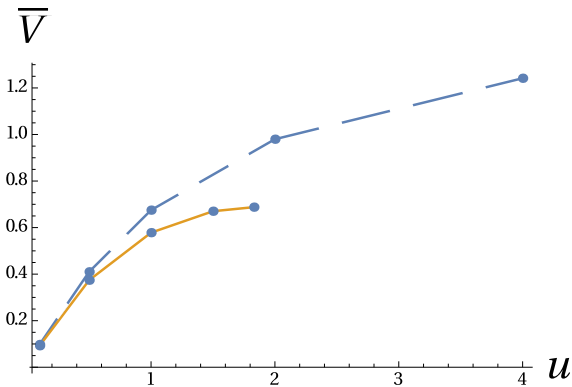}
\end{center}
\caption{\small $\bar{V}\equiv4\pi V$ vs deformation parameter $u$ at the sting length $L=4.5$.}
 \label{V-u}
\end{figure}

As can be seen from the  graphs in Figs.~2, 3 no solution with lower energy   exists for 
sufficiently small $L$  and/or high enough value of the deformation parameter. To explore 
this issue we need to find approximate analytical solutions.


\section{\boldmath$Z_{2N}$ broken phase}
\setcounter{equation}{0}
\label{zNbroken}

Consider first the large-$L$ domain or, more precisely,
 $L\gg 1/\Lambda$. In addition we assume that $u$ is not 
very large.
Then we use the second and third 
equations in (\ref{SPs}) to find the expressions
for  masses. Next,  we use (\ref{Vs}) to find  
 the vacuum energy. 

 We will show below that  
in the limit of large  $L\Lambda \gg 1$ and intermediate $u$ we have  $Lm_{b,f}\gg1$. If so, to 
find the boson and fermion masses  we can apply
the asymptotic behavior of the modified Bessel functions, 
\beq
K_0(z)\approx K_1(z)\approx\sqrt{\frac{\pi}{2z}}e^{-z}\,.
\eeq
Assuming that $LA_1=\pi$ we arrive at the following expressions for  masses:
\beqn
m_{b}^2&\approx&\Lambda^2\left(1 - \sqrt{\frac{8\pi}{\Lambda L}}e^{-\Lambda 
L}\right)\,,\nonumber \\
m_{f}^2&\approx&\Lambda^2e^{-u}\,\left\{1- \sqrt{\frac{8\pi}{\Lambda L}}
e^{\frac{u}{4}}e^{-\Lambda Le^{-u/2}}\right\}\,.
\label{mlarge}
\eeqn
If $L$ is large,  $L\Lambda \gg 1$, and the value of $u$ is neither too large nor too small, 
exponential corrections are small and
$m_{b,f}$ are of order of $\Lambda$. This justifies our approximation. As was already mentioned, 
$m_b$ and $m_f$ have a meaning of masses for bosons  $n^l$ and fermions $\xi^l$. Thus we have 
a non-vanishing mass gap in this phase.

\vspace{1mm}

From (\ref{mlarge})
we find  VEVs of $D$ and $\sigma$,
\beqn
iD &\approx& \Lambda^2\left\{1-e^{-u}-\sqrt{\frac{8\pi}{\Lambda L}}\left(
e^{-\Lambda L}-e^{-3u/4}e^{-\Lambda L e^{-u/2}}\right)\right\}\,,
\nonumber \\[3mm]
\sqrt{2}\sigma &\approx&\Lambda \,e^{-\frac{u}{2}}\,\left\{1- \sqrt{\frac{2\pi}
{\Lambda L}}e^{\frac{u}{4}}e^{-\Lambda Le^{-u/2}}\right\}\,e^{\frac{2\pi i k}{N}}\,,
\label{vevlargeL}
\eeqn
where $k=0, ..., (N-1)$.

The presence of 
non-zero $D$ signals that \ntwoo supersymmetry is spontaneously broken.
The vacuum energy is  
\beq
E\approx\frac{NL\Lambda^2}{4\pi}\left\{1-e^{-u}+\frac{2}{\Lambda L}
\sqrt{\frac{8\pi}{\Lambda L}}\left(e^{-\Lambda L}-e^{-u/4}e^{-\Lambda L e^{-u/2}}\right)\right\}\,.
\eeq

The phase of $\sigma$ in (\ref{vevlargeL}) is determined by the same phase factor as in (\ref{sigmavev}). 
We see that we have $N$ degenerative vacua, in much the same way as in the infinite volume
case. The degeneracy is  not due to supersymmetry but due to the fact that the discrete chiral $Z_{2N}$ 
symmetry is broken down to $Z_2$.

Our approximation assumes that both boson and fermion masses are large as compared to $1/L$. 
However, from (\ref{mlarge}) we see that $m_f$  exponentially decreases at large  $u$.
Our approximation breaks down when we increase $u$ above the curve 
\beq
L\Lambda \sim e^{\frac{u}{2}}.
\label{curve0}
\eeq

We will see in  Sec. \ref{zNunbroken} that in fact on this curve $\sigma $ becomes 
zero and the theory goes into $Z_{2N}$-symmetric phase.

\subsection{Quantum mechanics: the \boldmath$u\to 0$ limit }
\label{qm}

It was shown in  \cite{MSY15}  that the  VEV  of the $\sigma$ field in the  \cpn model 
with \ntwot super\-symmetry  does not depend on the string length. Since $L$
is not a holomorphic parameter \ntwot supersymmetry forbids the effective twisted superpotential
(which determines  the $\sigma$ VEV)  to depend  on $L$.

In the heterotic \cpn model   supersymmetry is spontaneously broken. Thus one 
can expect the $\sigma$ VEV  to depend on the string length. This is what we observe in 
equation (\ref{vevlargeL}). However, one can note 
that the expressions for the boson and fermion masses (\ref{mlarge}) in the limit of vanishing 
$u$ do  not reduce to those obtained in the \cpn model with \ntwot supersymmetry. It depends
on the string length even if $u=0$. What is happening?

To resolve this puzzle in this section we note that the $u\to 0$  limit turns out to be in conflict
with the quasiclassical approximation in the one-loop effective action which we use
 in the large-$N$ analysis.
We will see below that the relevant parameter is $uN^2$. Thus, the change of regime we expect to detect
occurs at $u\sim 1/N^2$ and is not seen in the standard treatment.
We must remember that the value of $LA_1$ is in its turn determined by a quantum-mechanical problem.
In other words, we must take into consideration fluctuations of this quantal variable. 

To detect this change of regimes we must
 consider a quantum-mechanical problem for the Polyakov line (\ref{Polyakovloop}) and 
average operators $\cos(LkA_1)$ that appear in the equations defining 
masses
(\ref{SPs}) over the ground state wave function. The equations for the masses
in the small-$u$ limit become
\beqn
\ln\frac{m_b^2}{\Lambda^2}&=&4\sum_{k=1}^\infty K_0(Lm_bk)\chi_k\,,\nonumber \\
\ln\frac{m_f^2}{\Lambda^2}&=&4\sum_{k=1}^\infty K_0(Lm_fk)\chi_k-u\,.
\eeqn
where the $\chi_k$ is the average value of the operator  $\cos(LkA_1)$ defined as
\beq
\chi_k=\int_{-\pi}^{\pi}L\,dA_1|\psi|^2\cos(LkA_1)\,.
\eeq
Here $\psi$ is the ground state wave function in quantum mechanics for $LA_1$.

In this way we obtain the masses
\beqn
m_{b\pi}^2&\approx&\Lambda^2\left(1+\sqrt{\frac{8\pi}{\Lambda L}}e^{-\Lambda 
L}\chi_1\right)\,,\nonumber \\
m_{f\pi}^2&\approx&\Lambda^2\left(1+\sqrt{\frac{8\pi}{\Lambda L}}e^{-\Lambda L}
\left(1+\frac{u\Lambda L}{2}-\frac{3u}{4}\right)\chi_1-u\right)\,,
\eeqn
where we expand the expressions for masses $m_b$ and $m_f$ at large $L$ and small $u$.
This expressions imply a smooth \ntwot limit if $\chi_1$ vanishes with $u$.

From equation (\ref{Vs}) one can read off the action for the $A_1$ quantal variable,
\beqn
S=\int dt \Bigg[\frac{L\dot{A^2_1}}{4e^2}&+&\frac{LN}{4\pi}\Bigg(8m_f^2\sum_{k=1}^
\infty\frac{K_1(Lm_fk)}{Lm_fk}\cos(LkA_1)\nonumber \\[3mm]
&&-8m_b^2\sum_{k=1}^\infty\frac{K_1(Lm_bk)}
{Lm_bk}\cos(LkA_1)\Bigg)\Bigg]\,.
\eeqn
In the large-$L$ limit the equation for the wave function is given by
\beq
\frac{d^2\psi}{d\phi^2}+(\lambda-2q\cos(2\phi))\psi=0\,,
\eeq
where $\phi=LA_1/2$, and the parameter $q$ is defined as follows:
\beq
q=\frac{uN^2e^{-\Lambda L}}{(2\pi\Lambda L)^{3/2}}\Lambda L\,,
\eeq
(please, observe its explicit dependence on $uN^2$).
This is the Mathieu equation. The solution for the wave function can be found 
numerically. The averaged value of $\cos(LA_1)$ is 
\beqn
\chi_1&=&-0.99 \quad \mbox{at} \quad \Lambda L=5 \quad \mbox{and} \quad uN^2=10^9 \nonumber \\[2mm]
\chi_1&=&-0.85 \quad \mbox{at} \quad \Lambda L=5 \quad \mbox{and} \quad uN^2=10^5\nonumber \\[2mm]
\chi_1&=&-10^{-3} \quad \mbox{at} \quad \Lambda L=5 \quad \mbox{and} \quad uN^2=10^1\,.
\eeqn
Thus we see that for large values of the deformation parameter the averaging plays almost no role, 
and the saddle point approximation works well. However, as the deformation parameter
gets smaller the averaged value of cosine
vanishes and the expression for fermion mass reduces to that obtained in the 
 \ntwot model.

A more transparent albeit qualitative analysis can be carried out if we use the harmonic 
oscillator approximation in our quantal problem. Then one can find the averaged value of 
$\cos LA_1$ analytically,
\beq
\chi_1\approx-\sqrt{uN^2 e^{-\Lambda L}}\left(\frac{2\pi}{\Lambda L}\right)^{1/4}\,.
\eeq
This result explicitly demonstrates vanishing of $\chi_1$ as the deformation parameter $uN^2$
tends to zero.
Thus we see that in the $u\rightarrow0$ limit  the solution of the \ntwoo model
tends to that of the  \ntwot model in the interval $u\in [0, \,{\rm const}/N^2]$. 

\section{The \boldmath{$Z_{2N}$} unbroken phase}
\setcounter{equation}{0}
\label{zNunbroken}

\begin{figure}[h!]
\begin{center}
\includegraphics[width=7.5cm]{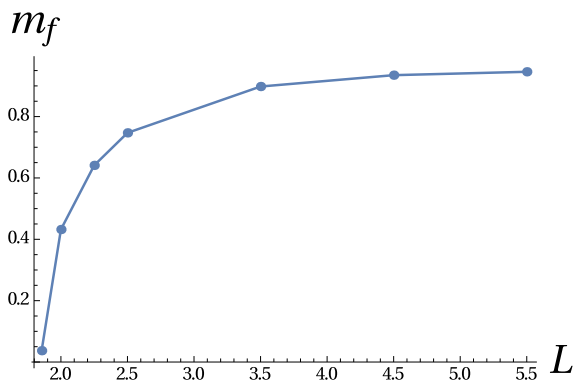}
\end{center}
\caption{\small Fermion mass $m_f$ vs string length $L$ at the value of the deformation parameter $u=0.1$.}
 \label{mf-L}
 \end{figure}

\begin{figure}[h!]
 \begin{center}
\includegraphics[width=7.5cm]{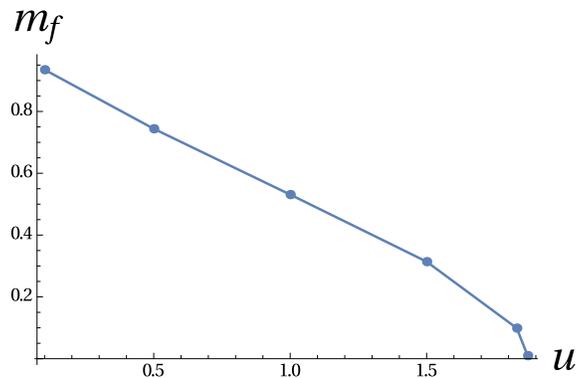}
\caption{\small Fermion mass $m_f$ vs deformation parameter $u$ at  $L=4.5$.}
\end{center}
 \label{mf-u}
\end{figure}

Now let us consider the region where $u$ is large, i.e. $u\gg \log \Lambda L$, see Eq. (\ref{curve0}).
For the time being we  assume that $L$ is still large, $L\gg 1/\Lambda$.
We can find approximate analytic solution 
for a curve in the ($L$, $u$) plane at which the $Z_{2N}$ broken phase with $N$ distinct vacua
ceases to exist (see the
phase diagram in Fig. 1). This  phase is terminated when the fermion mass (it is
always smaller or equal to the  boson mass) reaches zero as we increase $u$. 
Assuming that the fermion mass
is close to zero so that $Lm_{f}\ll 1$ we can approximate the sums of the Bessel 
functions in (\ref{SPs}). Noting that $\cos(\pi k)=(-1)^k$ we use (\ref{K0}) 
with $y=0$ to obtain the following expression for the fermion mass
\beq
(Lm_{f})^2S_2 \approx S_1+\gamma-\ln\frac{4\pi}{\Lambda L}-\frac{u}{2}\,,
\label{mfforL=pi}
\eeq
where $S_{1,2}$ are defined in (A.3).
Thus, the solution with non-zero $m_f$  exists only below the curve
\beq
\Lambda L\approx 4\pi e^{u/2-S_1-\gamma}\,.
\label{phasetransline}
\eeq
This formula gives a more accurate prediction for the curve (\ref{curve0}) which was obtained in the 
previous section.
Moreover, the minimal string length is $\Lambda L\approx1.76$. 
Numerical calculation also shows that the fermionic mass goes to zero at finite 
values of both  $L$ and $u$, as can be seen from Fig. (\ref{mf-L}) and (\ref{mf-u}). 

\begin{figure}[h!]
  \begin{center}
\includegraphics[width=7.5cm]{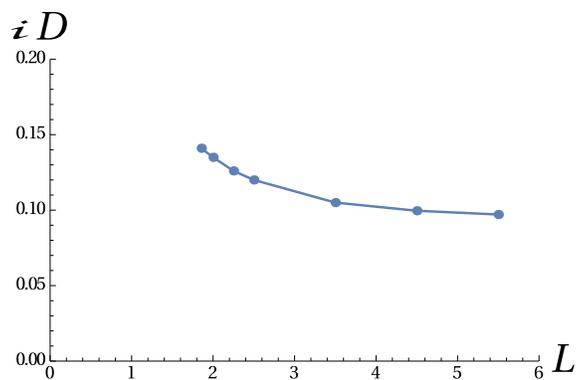}
\caption{\small $iD$ vs  $L$ at the value of the deformation parameter $u=0.1$.}
 \label{iD-L}
\end{center}
\end{figure}
\begin{figure}[h!]
 \begin{center}
\includegraphics[width=7.5cm]{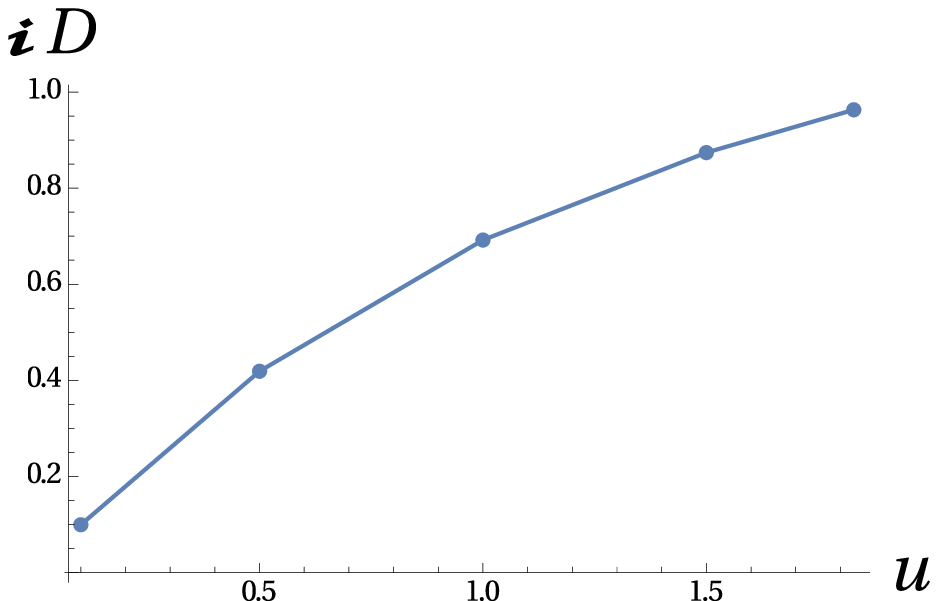}
\end{center}
\caption{\small $iD$ vs the deformation parameter $u$ at  $L=4.5$.}
\label{iD-u}
\end{figure}

Moreover it is clear from Figs. \ref{mf-L} and \ref{iD-L} that as 
$L\gg 1/\Lambda$ the fermionic mass $m_f$ tends to $\Lambda e^{-u}$ while $iD$ tends to
$\Lambda^2(1-e^{-u})$, in agreement with (\ref{mlarge}) and (\ref{vevlargeL}), 
respectively. One can also note that $iD\rightarrow 0$ as $u\rightarrow 0$. This 
is expected since the $u=0$ limit  corresponds to the  \ntwot model.

Above  the  curve (\ref{phasetransline}), the only solution of the second equation in (\ref{SPs}) is
\beq
\sigma =0\,,
\label{sigma0}
\eeq
while  the boson mass 
\beq
m_{b}^2\approx\Lambda^2\left(1 - \sqrt{\frac{8\pi}{\Lambda L}}e^{-\Lambda 
L}\right)\,
\label{mb}
\eeq
is still given by the same expression as in the $Z_{2N}$ broken phase, 
see (\ref{mlarge}).

Note that the {$Z_{2N}$} unbroken phase we have observed is 
quite remarkable. On the phase transition line
$N$ vacua fuse to one, a family of split {$Z_{2N}$}-symmetric vacua does not emerge. 
We will discuss this circumstance later.

 \subsection{The L\"uscher term.}

Using the expression (\ref{K-1cos}) from Appendix we find that the  vacuum energy in this phase
is independent on $u$ and given by
\beq
E\approx\frac{LN\Lambda^2}{4\pi}
\left(1+\frac{2}{\Lambda L}\sqrt{\frac{8\pi}{\Lambda L}}e^{-\Lambda L}\right)-\frac{\pi N}{6L}\,.
\label{Evacsigma0}
\eeq
The second term here  is the L\"uscher term. It arises 
due to massless fermions. 
Note, that it equals to
half of what we found for non-supersymmetric theory \cite{MSY15} where it comes from bosons. 
The reason is that now the gauge holonomy is non-trivial, $A_1=\pi/L$.
Moreover, the same reason ensures that although the L\"uscher term in (\ref{Evacsigma0})  
 comes from fermions it still gives negative contribution to the energy
as compared to the non-supersymmetric case.

The vacuum energy  (\ref{Evacsigma0}) can be compared to the vacuum energy in the $Z_{2N}$ broken 
phase below the curve (\ref{phasetransline}) in the limit of 
$Lm_f\ll1$,
\beq
E\approx\frac{LN\Lambda^2}{4\pi}\left(1+\frac{2}{\Lambda L}\sqrt{\frac{8\pi}{\Lambda L
}}e^{-\Lambda L}\right)-\frac{\pi N}{6L}-\frac{NS_2}{4\pi L}(Lm_f)^4\,.
\eeq
The energy difference is approximately given by the last term above. Equation (\ref{mfforL=pi}) tells us that
the energy difference behaves as $\sim(L-L_c(u))^2$ near the phase transition curve, where $L_c(u)$ 
is given by 
(\ref{phasetransline}).

In summary,  we conclude that as we increase $u$ and cross the curve (\ref{phasetransline}) our
system goes through a line of {\it third} order phase transitions into the phase with $\sigma =0$.
All $N$ vacua coalesce in the $\sigma$ plane and $Z_{2N}$ symmetry is restored. 
In the infrared  limit our theory in this phase flows to a conformal limit which is a free theory
of massless fermions $\xi^l$.

\subsection{What happens with the \boldmath$A_\mu$ auxiliary field
in the {$Z_{2N}$} unbroken phase}

As we move  into
the {$Z_{2N}$} unbroken phase by increasing $u$ we could, in principle, observe two distinct scenarios:
 the $N$ former vacua of the $Z_{2N}$ broken phase  which fuse themselves into $\sigma=0$ in phase III, in fact, split in energy, with
$N-1$ of them becoming quasivacua,
and only one of them remaining as the true vacuum. This phase would be quite similar to the 
Coulomb/confinement
phase in the non-supersymmetric \cpn model \cite{W,MSY15}. 

The second option is to have just a unique vacuum at $\sigma=0$, with no accompanying family of quasivacua.
One can decide between the two options by analyzing the  auxiliary field $A_\mu$.

\begin{figure}[h!]
\centering
\includegraphics[width=14cm]{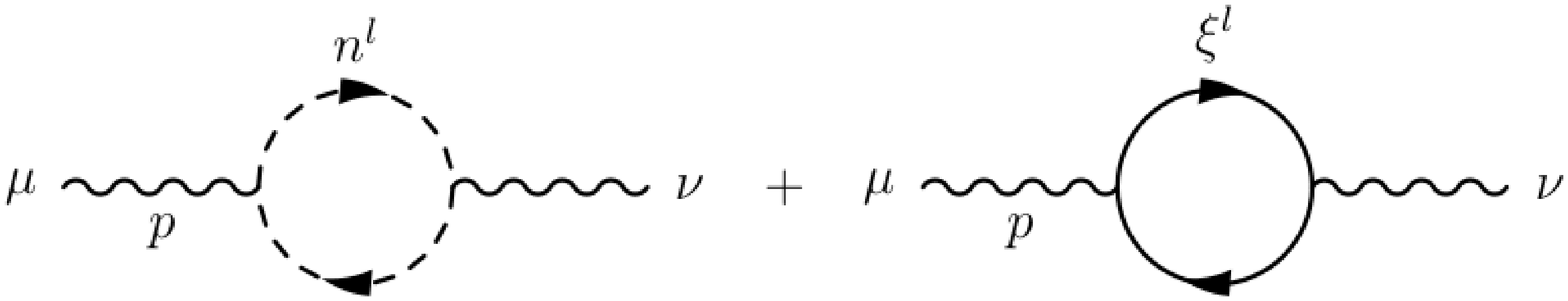}
\caption{One loop diagrams that contribute to the photon kinetic term.}
\label{photon_kin}
\end{figure}

We need to evaluate the two diagrams shown in Fig. (\ref{photon_kin}).
The first diagram comes from bosons $n^l$. In much the same way as in the non-supersymmetric
CP$(N-1)$ model it produces a kinetic term for the photon in the Lagrangian, 
\beq
\frac{1}{4e^2}F_{kl}^2\,,
\eeq
where at  large $L$ the expression for the 
charge $e^2$ is given by 
\beq
\frac{1}{e^2}\approx\frac{N}{12\pi\Lambda^2}\,.
\eeq

This makes U(1) gauge field dynamical \cite{W}. In the non-supersymmetric model this leads to 
confinement of electric charges. The reason is that  the static Coulomb potential in two dimensions 
is linear and ensures that the charged $n^l$ states are linearly confined in the non-supersymmetric model \cite{W}.
Similar Coulomb/confining phase occur in the compactified non-supersymmetric CP$(N-1)$ model at 
large $L$ \cite{MSY15}. Confinement of $n^l$ states can be interpreted as a small split between quasivacua
involved in the $\theta$-angle evolution \cite{W2,GSY1}. In this picture the $n^l$ states  are interpreted as 
kinks interpolating between
true vacuum an the first quasivacuum.

On the other hand, in our \ntwoo theory  we have also the  second diagram coming from massless fermions.
It produces a mass term for the  photon
\beq
V(A_1) = \frac{N}{2\pi}\left(A_1-\frac{\pi}{L}\right)^2\,.
\eeq
Evaluation of the coefficient $N/2\pi$ is presented in Appendix B. This term is a manifestation of the 
chiral anomaly and appears in much the same way as in the Schwinger model. 

Therefore, the 
photon obtains a mass
\beq
m_\gamma\approx\sqrt{12}\Lambda\,.
\eeq
The photon mass ensures the exponential fall-off of the  electric potential between charged
sources. Thus, there is no confinement in the $\sigma =0$ phase of our  (0,2)  supersymmetric \cpn model.

This ensures the absence of  fine vacuum structure with split quasivacua. In fact there is no $\theta$ dependence
in the theory with massless fermions, and the argument of \cite{W2} does not apply. We have a single
vacuum with the unbroken $Z_{2N}$ symmetry and no family of quasivacua in  the $\sigma =0$ phase (i.e. phase III
in Fig. 1).
 This is a new phase in the \cpn model which was not known before.

\section{Broken {\boldmath\sun} symmetry phase}
\setcounter{equation}{0}
\label{brphase}

Now let us consider the region of small $L$.
At small $L$ the theory enters a weak coupling regime so we expect the emergence of the classical
 picture in the limit $N\to\infty$. Classically CP$(N-1)$ model has $2(N-1)$ massless states 
which can be viewed as Goldstone states
of the broken SU$(N)$ symmetry. To study this possibility much in the same way as in \cite{MSY15,GSYphtr} 
we assume that one component of the field
$n^l$, say  $n^1\equiv n$ can develop VEV and we integrate over all other components 
of $n^l$ 
in the external fields  $n$, $\sigma$ $D$ and $A_1$.
However now in order not
to break supersymmetry by the boundary conditions we have to leave out one 
component of $\xi$ fields as well. Due to the constraint (\ref{orth}) we can choose 
these components to be $\xi_{L,R}^N\equiv\xi_{L,R}$. The expression for the energy
is  
\beqn
E&=&\frac{LN}{4\pi}\Bigg[iD-iD\ln\frac{m_b^2}{\Lambda^2}-m_f^2\ln\frac
{m_b^2}{m_f^2}+m_f^2u\nonumber \\[2mm]
&+&8m_f^2\sum_{k=1}^\infty\frac{K_1(Lm_fk)}{Lm_fk}\cos(kLA_1)-8m_b^2\sum_{k=1}^
\infty\frac{K_1(Lm_bk)}{Lm_bk}\cos(kLA_1)\Bigg]\nonumber \\[2mm]
&+&L\left[(m_b^2+A_1^2)|n|^2+i\sqrt{2}\sigma\bar\xi_R\xi_L+i\sqrt{2}\sigma^\star
\bar\xi_L\xi_R\right]
\nonumber \\[2mm]
&+& i\bar\xi_L\xi_LLA_1-i\bar\xi_R\xi_RLA_1\nonumber \\[2mm]
&+&N\left[\sqrt{m_f^2+A_1^2}-\sqrt{m_b^2+A_1^2}\right]\,,
\label{Vb}
\eeqn
where the first two lines are the same as in (\ref{Vs}), the third and fourth lines correspond
to components which we left out of integration, and the last line gives the
contribution due to omission of  the  zero modes.

\subsection{Saddle point approximation}

Proceeding as in the \sun symmetric case we obtain the following set equations that
defines a stationary point
\beqn
0&=&(m_b^2+A_1^2)n\,,
\label{dern}\\[3mm]
0&=&\sqrt{2}\sigma\xi_L-\xi_RA_1=\sigma^\star\xi_R+\xi_LA_1\,,
\label{derxi}\\[3mm]
|n|^2
\!\!\!&=&\!\!\!
\frac{N}{L}\Big[\frac{1}{2\sqrt{m_b^2+A_1^2}}+\frac{L}{4\pi}\ln\frac{m_b^2}
{\Lambda^2}-\frac{L}{\pi}\sum_{k=1}^\infty K_0(Lm_bk)\cos(kLA_1)\Big],
\label{derD}\\[2mm]
0
\!\!\!&=&\!\!\!
N\Big[\frac{2Lm_b}{\pi}\sum_{k=1}^\infty K_1(Lm_bk)\sin(kLA_1)-\frac{2Lm_f}{\pi}
\sum_{k=1}^\infty K_1(Lm_fk)\sin(kLA_1),\nonumber \\
&-&\frac{A_1}{\sqrt{m_b^2+A_1^2}}+\frac{A_1}{\sqrt{m_f^2+A_1^2}}\Big]+2LA_1|n|^2
+iL\bar\xi_L\xi_L-iL\bar\xi_R\xi_R
\label{derA}\\[2mm]
0\!\!\!
&=&
\!\!\!Li\sqrt{2}\bar\xi_L\xi_R+2\sigma\Big[L|n|^2+N\Big(-\frac{1}
{2\sqrt{m_b^2+A_1^2}}+\frac{1}{2\sqrt{m_f^2+A_1^2}} \nonumber \\[2mm]
&+&\frac{L}{\pi}\sum_{k=1}^\infty K_0(Lm_bk)\cos(kLA_1)
-\frac{L}{\pi}\sum_{k=1}^\infty K_0(Lm_fk)\cos(kLA_1)\Big),\nonumber \\[3mm]
&+&\frac{LN}{4\pi}\Big(u-\ln\frac{m_b^2}{m_f^2}\Big)\Big]\,.
\label{dersigma}
\eeqn
From (\ref{dern}) we conclude that $m_b=A_1=0$. Then (\ref{derA}) does not have a 
solution unless $\sigma=0$. We also see that $\bar{\xi}_{L,R}=\xi_{L,R}=0$ satisfies
the above system of equations. We find that $n^l$ field develops a vacuum expectation value 
\beq
|n|^2=\frac{N}{2\pi}\left(\ln\frac{4\pi}{\Lambda L}-\gamma\right)\,,
\eeq
which implies in turn that this solution exists only for $\Lambda L\le 7.05$. The energy is
found to be zero as in the supersymmetric case, see phase \mbox{I} in Fig. (\ref{phasediag}).

This phase is similar to the dynamical regime  we found previously in the non-supersymmetric \cpn model \cite{MSY15}.
In particular, the VEV of $n^l$ breaks global \sun symmetry implying the presence of 
$2(N-1)$ real massless degrees of freedom. As we already mentioned the dynamics of 
the \cpn model in this phase is
determined by quasiclassical approximation in the action (\ref{Sb}). At small $L$ the theory 
is at weak coupling  because the inverse coupling
constant $r$ is determined by
\beq
r= \frac{N}{2\pi}\log{\frac{1}{L\Lambda}}.
\eeq
The constant $r$ grows large at small $L$.

 However, we do not expect exactly
massless modes to appear in $1+1$ dimensions because of Coleman's theorem \cite{C}. We 
found the above solution in the leading order in $1/N$. It holds only in the limit $N=\infty$. 
Thus, we should expect higher order corrections to modify the result. In 
particular, the would-be Goldstone massless modes may acquire  small  masses suppressed in the large 
$N$ limit. 
As a consequence the energy might
 be uplifted from zero. 

The solution that we found is completely $u$-independent.
Thus we expect that the vacuum energy in the broken phase is given by $E_{br}$ which is independent on $u$ 
and suppressed at large $N$.

\section{Quantum mechanics at small \boldmath$L$:\\[1mm]
 $u\to 0$ limit}
\setcounter{equation}{0}
\label{qmsmallL}

Now we have to study the limit $u\to 0$ at small $L$ where the theory should match the
\ntwot \cpn model which has a single \sun symmetric ($Z_{2N}$ broken) phase with the mass gap independent of
$L$. Clearly \sun broken phase cannot explain this limit because it is $u$-independent.
Our analysis in this section has a qualitative nature.  As we have already seen, for the case of 
large $L$ the transition occurs at  $uN^2\sim 1$ where the large-$N$ approximation strictly speaking 
is not applicable. 

\begin{figure}[h!]
\begin{center}
\includegraphics[width=7.5cm]{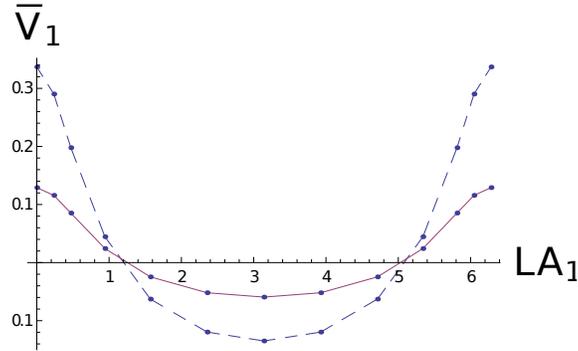}
\end{center}
\caption{\small Dependence of potential 
$\displaystyle\bar{V}_1\equiv\frac{\pi L^2}{2N}V$ on the deformation parameter $u$.}
 \label{V1-A}
\end{figure}

Below  we argue that the \sun symmetric phase reappear again when we go to the limit of extremely small
$u$ keeping $L$ small, $L\ll 1/\Lambda$. Assuming that both $Lm_{b,f}\ll1$ in this phase we use 
(\ref{K-1cos}) to find the expression for the potential valid for $LA_1$ close to 
$\pi$
\beq
V(\tilde{A}_1)\approx\frac{NL^2}{\pi}\tilde{A}_1^2\left(m_b^2-m_f^2\right)S_2\,,
\eeq
where $\tilde{A}_1\equiv A_1-\pi/L$. By analogy with (\ref{mfforL=pi}) one can find 
the expression for the bosonic mass 
\beq
(Lm_{b})^2S_2 \approx S_1+\gamma-\ln\frac{4\pi}{\Lambda L}\,.
\eeq
Thus the expression for the potential is given by
\beq
V(\tilde{A}_1)\approx\frac{Nu}{2\pi}\tilde{A}_1^2\,,
\eeq
Hence, as $u$ gets smaller the potential becomes weaker and flatter. When $LA_1$ 
gets close to $0$ or $2\pi$ the above expression becomes invalid. The results of 
numerical calculations are given in Fig. (\ref{V1-A}). Two curves correspond to two
values of deformation parameter $u=0.05$ and $u=0.1$ (dashed curve). One can see that 
the expression we derived is in a good agreement with numerical results. As $u$ gets 
smaller the amplitude of the potential also decreases. 

We see that in the limit $u\to 0$ the potential $V(\tilde{A}_1)$ becomes flat and we have to 
average over $A_1$ (instead of taking the saddle point value $A_1=\pi/L$) in much the similar 
way as we did in Sec.~\ref{qm} for  the region of large $L$.
The averaging procedure gives us \ntwot limit.

More exactly the  vacuum energy in \sun symmetric  phase at  extremely small $u$ is given by
\beq
E_{sym}\approx \frac{uN}{4\pi}\Lambda^2\,L\,.
\label{EvacQM}
\eeq

 Comparing this with the vacuum energy $E_{br}$ in the \sun broken phase which is independent of  $u$ 
 we see that at  very small critical $u_c\sim 1/N^2$,
the energy in the \sun unbroken phase  becomes lower then that  in the \sun broken phase, 
and the system
undergoes a phase transition into \sun symmetric phase, see  Fig.~1. The \sun symmetric  phase has a 
perfectly smooth $u\to 0$ limit.

\section{Conclusions}
\label{conclusions}

To summarize, we find three different phases, see Fig 1. At large $L$ and intermediate 
values of the deformation parameter $u$  there is a phase with a mass gap, $N$ vacua and  
broken discrete $Z_{2N}$ symmetry. As we increase $u$ a reach a critical value (which grows with $L$)
 we find a phase transition to the $Z_{2N}$ symmetric phase, 
with a unique vacuum.
 The line separating these two \sun symmetric phases is a line of a third 
order phase transitions in the large $N$ limit.

 As the string under consideration gets  shorter we find a phase transition to 
a phase with the broken \sun symmetry (phase II). In this phase we expect masses of the $n$ fields to be much 
smaller than in two \sun symmetric  phases. In fact, at $N=\infty$ they vanish. At small $L$ and extremely small $u$
we expect another  phase transition from the  \sun broken phase into the  \sun unbroken phase which has
a smooth $u\to 0$ limit.

Strictly speaking, our description of the underlying dynamics in terms of the phase transitions is valid only at 
$N=\infty$.
At large but finite $N$ one can expect that all phase transitions become rapid crossovers.

\section*{Acknowledgments}

This work  is supported in part by DOE grant DE-SC0011842. 
The work of A.Y. was  supported by William I. Fine Theoretical Physics Institute  of the  University of 
Minnesota, and by Russian State Grant for
Scientific Schools RSGSS-657512010.2. The work of A.Y. was supported by Russian Scientific Foundation 
under Grant No. 14-22-00281.

\section*{Appendix A:\\Relations for modified Bessel functions}

\renewcommand{\theequation}{A.\arabic{equation}}
\setcounter{equation}{0}

In this Appendix we derive all the relations for the sums of modified Bessel 
functions of the second kind used in the text. We will use the following asymptotic
behavior 
\beq
K_1(z)\rightarrow\frac{1}{z}\quad\mbox{as}\quad z\rightarrow0\,,
\label{Kz0}
\eeq
as well as the properties of derivatives 
\beq
K_0(z)^\prime=-K_1(z)\;\;\;\mbox{and}\;\;\;K_1^\prime(z)=-K_0(z)-\frac{K_1(z)}{z}\,,
\label{Kprime}
\eeq
and the following approximations, valid to order $O(y^2,z^2)$ (see formula 8.526 in \cite{GrRY})
\beqn
\sum_{k=1}^\infty K_0(zk) \cos(yk)&=&\frac{\gamma}{2}+\frac{1}{2}\ln\frac{z}{4\pi}+
\frac{\pi}{2\sqrt{z^2+y^2}}+S_0(2y^2-z^2)+\delta_0\,,\nonumber \\
\sum_{k=1}^\infty K_0(zk)(-1)^k\cos(yk)&=&\frac{\gamma}{2}+\frac{1}{2}\ln\frac{z}{4\pi}+
\frac{S_1}{2}+\frac{S_2}{2}(2y^2-z^2)+\delta_1\,,
\label{K0}
\eeqn
where $\delta_{0,1}\sim y^2z^2$ and we defined the sums
\beqn
S_0&=&\sum_{l=1}^\infty \frac{\pi}{(2\pi l)^3}\approx0.015\,,\quad 
S_1=\sum_{l=1}^\infty \frac{1}{l(2l-1)}\approx1.386\,,\nonumber \\
S_2&=&\sum_{l=1}^\infty \frac{1}{\pi^2(2l-1)^3}\approx0.107\,.
\eeqn

To find the sum involving cosine we notice that on one hand
\beq
\frac{d}{dz}\left(z\sum_{k=1}^\infty\frac{K_1(zk)}{k}\cos(yk)\right)=
-z\sum_{k=1}^\infty K_0(zk)\cos(yk)\,,
\label{K1z}
\eeq
and on the other hand 
\beq
\frac{d}{dy}\left(\sum_{k=1}^\infty\frac{K_1(zk)}{k}\cos(yk)\right)=
-\sum_{k=1}^\infty K_1(zk)\sin(yk)\,,
\label{K1y}
\eeq
moreover the following relation also holds
\beq
\frac{d}{dz}\left(\sum_{k=1}^\infty K_0(zk)\cos(yk)\right)=-
\frac{d}{dy}\left(\sum_{k=1}^\infty K_1(zk)\sin(yk)\right)\,,
\label{K01}
\eeq
where we used (\ref{Kprime}) several times.

First using (\ref{K1z}) and the expansion from (\ref{K0}) we find to order 
$O(y^2,z^2)$
\beqn
\sum_{k=1}^\infty\frac{K_1(zk)}{k}\cos(yk)&\approx&-\frac{\pi\sqrt{z^2+y^2}}{2z}-
\frac{z(2\gamma-1)}{8}-\frac{z}{4}\ln\frac{z}{4\pi}\nonumber \\
&-&S_0zy^2+\frac{f_1(y)}{z}
\eeqn
where $f_1(y)$ depends on $y$.

Now using (\ref{K01}) and approximation (\ref{K0}) we find that 
\beq
\sum_{k=1}^\infty K_1(zk)\sin(yk)\approx\frac{\pi y}{2z\sqrt{z^2+y^2}}-
\frac{y}{2z}+2S_0zy+f_2(z)\,,
\eeq
where $f_2(z)$ is a function which depends on $z$. Since LHS vanishes when 
$y=0$ and $z\neq0$ we conclude that $f_2(z)=0$. Now from (\ref{K1y}) we find that 
\beq
\sum_{k=1}^\infty\frac{K_1(zk)}{k}\cos(yk)\approx-\frac{\pi\sqrt{z^2+y^2}}{2z}+
\frac{y^2}{4z}-S_0zy^2+f_3(z)\,,
\eeq
where $f_3(z)$ depends on $z$. 

To fix $f_1(y)$ and $f_3(z)$ we use the property (\ref{Kz0})
 and find that 
\beq
\sum_{k=1}^\infty\frac{K_1(zk)}{k}\cos(yk)\rightarrow\sum_{k=1}^\infty
\frac{\cos(yk)}{zk^2}=\frac{1}{z}\left(\frac{y^2}{4}-\frac{\pi y}{2}+\frac{\pi^2}
{6}\right)\,.
\eeq
Thus we conclude that 
\beqn
\sum_{k=1}^\infty\frac{K_1(zk)}{k}\cos(yk)&\approx&-\frac{\pi\sqrt{z^2+y^2}}{2z}+
\frac{y^2}{4z}+\frac{\pi^2}{6z}-S_0zy^2\nonumber \\
&-&\frac{z(2\gamma-1)}{8}-\frac{z}{4}\ln\frac{z}{4\pi}\,.
\label{K1cos}
\eeqn
In a similar way we find that 
\beqn
\sum_{k=1}^\infty\frac{K_1(zk)}{k}(-1)^k\cos(yk)&\approx&-\frac{z(2S_1+2\gamma-1)}{8}-\frac{z}{4}\ln\frac{z}{4\pi}\nonumber \\
&-&\frac{\pi^2}{12z}+\frac{y^2}{4z}-\frac{S_2}{2}zy^2\,.
\label{K-1cos}
\eeqn

\section*{Appendix B:\\Photon mass}
\renewcommand{\theequation}{B.\arabic{equation}}
\setcounter{equation}{0}

In this Appendix we derive an expression for the photon mass. Due to gauge 
invariance both the diagrams in Fig. (\ref{photon_kin}) have to be of the form
\beq
\Pi_{ij}=\Pi(p^2)\left(p^2\delta_{ij}-p_ip_j\right)\,.
\eeq
Below we show that for the second diagram $\Pi(p^2)$ has a pole which means that 
photons acquire mass. We put $p_1=0$ and evaluate $\Pi_{11}$:
\beqn
\Pi_{11}&=&-\frac{1}{L}\Big[\sum_{k=-\infty}^\infty\int_{-\infty}^\infty\frac{dq_0}
{2\pi}\frac{2q_1^2-2q_0(p_0+q_0)-2m_1^2}{(q_0^2+q_1^2+m_1^2)(p_0^2+2p_0q_0+q_0^2+
q_1^2+m_1^2)}
\nonumber\\[2mm]
&-&[m_1\leftrightarrow m_2]\Big]\,,
\eeqn
where $m_1$ is the fermion mass, which we put to zero at the end, $m_2$ is the mass
of Pauli-Villars regulator, and $q_1$ is a discrete momentum
\beq
q_1=\frac{2\pi k}{L}+A_1=\frac{\pi}{L}\left(2k+1\right)\,.
\eeq
We introduce Feynman parameter $x$ and substitute integration variable 
$q_0=l-p_0x$
\beq
\Pi_{11}=-\frac{1}{L}\Big[\sum_{k=-\infty}^\infty\int_{-\infty}^\infty\int_0^1\frac{dldx}
{2\pi}\frac{2q_1^2-2m_1^2+2p_0^2x(1-x)-2l^2}{[l^2+m_1^2+q_1^2+xp_0^2-x^2p_0^2]^2}-[m_1\leftrightarrow m_2]\Big]\,,
\eeq
where terms linear in $l$ drop out. Integrating over $l$ one finds
\beq
\Pi_{11}=\frac{1}{L}\Big[\sum_{k=-\infty}^\infty\int_0^1 dx \frac{m_1^2}{[m_1^2+
q_1^2+xp_0^2-x^2p_0^2]^{3/2}}-[m_1\leftrightarrow m_2]\Big]\,,
\eeq
and since $m_1=0$ the first term vanishes and only the contribution from the regulator
remains. To integrate over $x$ we use third Euler's substitution
\beq
\sqrt{-p_0^2x^2+p_0^2x+m^2+q_1^2}=\sqrt{-p_0^2(x-x_1)(x-x_2)}=t(x-x_1)\,.
\eeq
One can easily check that neither of the roots belong to the interval $x\in[0,1]$ and 
thus this substitution is justified. After integration we obtain the following sum
\beq
\Pi_{11}=-\frac{1}{L}\sum_{k=-\infty}^\infty\frac{m_2^2}{(q_1^2+m_2^2+\frac{p_0^2}
{4})\sqrt{q_1^2+m_2^2}}\approx-\frac{1}{L}\sum_{k=-\infty}^\infty\frac{m_2^2}
{(q_1^2+m_2^2)^{3/2}}\,,
\eeq
where we ignore $p_0$ compared to $m_2$. Evaluating this sum (see Appendix in 
\cite{MSY15}) we finally obtain (setting $m_2\rightarrow\infty$)
\beq
\Pi_{11}=-\frac{1}{\pi}\,,
\eeq
which tells us that $\Pi(p^2)$ indeed contains a pole
\beq
\Pi(p^2)=-\frac{1}{\pi p^2}
\eeq
and the photon becomes massive.

\end{document}